# Coumarin Dyes for Dye-Sensitized Solar Cells – A Long-Range-Corrected Density Functional Study

*Bryan M. Wong\* and Joseph G. Cordaro*

Materials Chemistry Department, Sandia National Laboratories, Livermore, California 94551

*Corresponding author. E-mail: bmwong@sandia.gov.

The excited-state properties in a series of coumarin solar cell dyes are investigated with a long-range-corrected (LC) functional which asymptotically incorporates Hartree-Fock exchange. Using time-dependent density functional theory (TDDFT), we calculate excitation energies, oscillator strengths, and excited-state dipole moments in each of the dyes as a function of the range-separation parameter, $\mu$. To investigate the acceptable range of $\mu$ and assess the quality of the LC-TDDFT formalism, an extensive comparison is made between LC-BLYP excitation energies and approximate coupled cluster singles and doubles (CC2) calculations. When using a properly-optimized value of $\mu$, we find that the LC technique provides a consistent picture of charge-transfer excitations as a function of molecular size. In contrast, we find that the widely-used B3LYP hybrid functional severely overestimates excited-state dipole moments and underestimates vertical excitations energies, especially for larger dye molecules. The results of the present study emphasize the importance of long-range exchange corrections in TDDFT for investigating the excited-state properties in solar cell dyes.

## I. Introduction

Dye-sensitized solar cells (DSCs) have gained immense interest in the last few years due to their potential for converting clean solar energy to electricity at low cost. Originally introduced by Grätzel and co-workers in 1991,[1] several researchers have studied their use in fabricating nanostructured films,[2-4] improving photovoltaic devices,[5-7] and a general scientific understanding of charge-transfer between material interfaces.[8-10] Although the origins of this work had focused on ruthenium dye complexes, current research is now directed towards organic dye sensitizers which are less expensive and easier to synthesize.[11] Recognizing this need for low-cost alternatives to conventional photovoltaics, Hara and co-workers[12-14] recently synthesized a series of coumarin dye molecules comprised of donor, electron-conducting, and anchoring groups for use in solar cell applications. As shown in Fig. 1, these organic sensitizers consist of an aniline group acting as an electron donor and a cyanoacrylic acid group functioning as an acceptor. These two functional groups are spatially separated from each other by one or more conjugated units which serve as an electron conductor. In particular, the NKX-2*xxx* series of courmarin derivatives have been reported as promising candidates for DSCs, and the NKX-2677 dye alone has a conversion efficiency of 7.4% which is comparable to ruthenium-based photosensitizers.[15-17]

The generation of a photocurrent in these DSCs occurs when a photon of sufficient energy is absorbed by the aniline moiety, resulting in an electron being promoted into an excited state (typically the singlet $S_1$ state). A simplified schematic of the energetic processes occurring in a conventional dye-sensitized solar cell is illustrated in Fig. 2. Upon excitation (1), the $\pi$-conjugated units facilitate an ultrafast flow of electrons from the aniline electron donor towards the cyanoacrylic acid acceptor. This intramolecular process is associated with the transfer of electrons into a highly-localized orbital which is energetically coupled with the conduction band of a $TiO_2$ semiconductor (2). After the electrons are injected into the $TiO_2$ anode, the oxidized solar cell dye is regenerated by electron transfer from $I_3^-$ ions in the electrolyte (3). The $I_3^-$ ions themselves are rapidly reduced to $I^-$ ions by a metal cathode (4), thus completing the regenerative cycle.

In order to develop highly efficient sensitizers, there are a few requirements which guide the design of molecular DSCs. First, the energy level of the $S_0$ ground state must lie below the redox couple of the $I_3^-$ ions so that the DSC dye can readily accept electrons from the energetically-higher electrolyte. Similarly, the excited $S_1$ state must be situated *above* the conduction band of the semiconductor so that electrons can migrate into the energetically-lower $TiO_2$ anode. Next, the $S_1$ excited state must possess significant charge-transfer character which can direct electron flow from the light-harvesting functional groups towards the semiconductor anode. Finally, to ensure high-efficiency, the dye should have a broad absorption spectrum in the visible region to absorb more photons from the sun.

Ideally, one would like to use a computational design beforehand to satisfy all of these requirements and predict the electronic properties which yield an efficient charge separation. An important area where theory should contribute is the prediction of DSC charge-transfer effects arising from different electron donor/acceptor groups in order to ultimately guide the organic synthesis. Unfortunately, despite the growing experimental efforts in DSC synthesis, the quantitative understanding and theoretical prediction of DSC charge-transfer properties is still a major hurdle to overcome in order to achieve a viable technology. While wavefunction-based approaches such as complete active space second-order perturbation theory (CASPT2) can yield accurate predictions, the computational cost of CASPT2 is too high for routine application to molecules larger than about 20 atoms. Consequently, most of the limited studies on DSC dyes have used time-dependent density functional theory (TDDFT) in conjunction with the popular B3LYP hybrid functional to study the optical response of these large molecules.[18-19] However, as shown by Kurashige and co-workers, the application of B3LYP to these large systems leads to a severe underestimation of excitation energies in the charge-transfer states.[20] This problem is further complicated by the fact that the performance of the B3LYP functional degrades with increased system size.[21-22]

Recognizing the shortcomings of conventional hybrid functionals, major methodological progress has been made in DFT techniques which incorporate a position-dependent admixture of Hartree Fock (HF) exchange in the exchange-correlation functional.[23-38] Originally developed by Gill[27] and Savin,[28,29,31] these range-separated functionals partially account for long-range charge-separation effects by adding a growing fraction of exact exchange when the interelectronic distance increases (see Sec.

II). This range-separation technique has been further modified and applied in many forms by Hirao *et al.* in their LC (long-range corrected) functional[23-26,37] and by Handy *et al.* with their CAM-B3LYP (Coulomb-attenuating method-B3LYP)[30,32,33] methods (as a side note, the abbreviation for "long-range corrected" is sometimes denoted as LRC instead of LC in the literature). There has also been recent work in constructing new LC functionals based on semilocal exchange hole approaches developed by the Scuseria group.[35,40,41] Their method is an alternative to the Hirao ansatz and also provides new routes for designing GGA exchange holes for range-separated hybrid functionals. Although the range-separation technique is based on a more physical model of the exchange potential, this approach has already shown some limitations. In particular, several investigators have demonstrated that the choice of the range-separation parameter, $\mu$, is strongly system-dependent.[42-47] Recently, Rohrdanz and Herbert have shown that the optimal choice of $\mu$ is considerably different for ground and excited-state properties even in the same molecular system.[43]

In this work, we test the performance of the LC-TDDFT technique proposed by Hirao *et al.*[23-26,37] on the excited-state energies and properties of DSCs. This computationally efficient approach has already been applied to several small molecular properties and a few large systems; however, no computations on DSC dyes using the LC-TDDFT approach have been reported yet. In the present study we apply LC-TDDFT to investigate the excited-state properties of the large DSC dye systems depicted in Fig. 1. In each case, we compute LC-TDDFT excitation energies, oscillator strengths, and $S_0/S_1$ dipole moments as a function of $\mu$ and compare them against approximate coupled-cluster singles and doubles (CC2) data. The importance of choosing the optimal range of $\mu$ and a comparison to predictions by the widely-used B3LYP functional[48] are also examined. Based on overall trends in excitation energies and properties, we find that the LC-TDDFT formalism significantly improves the description of excited-state properties of DSCs compared to conventional hybrid functionals.

**II. Theory and Methodology**

In the conventional LC scheme for DFT,[23-26,37] the electron repulsion operator $1/r_{12}$ is split into short- and long-range components using the standard error function, erf, as

$$\frac{1}{r_{12}} = \frac{1-\text{erf}(\mu r_{12})}{r_{12}} + \frac{\text{erf}(\mu r_{12})}{r_{12}}, \qquad (1)$$

where $r_{12} = |\mathbf{r}_1 - \mathbf{r}_2|$ is the interelectronic distance between electrons at coordinates $\mathbf{r}_1$ and $\mathbf{r}_2$, and $\mu$ is an adjustable damping parameter having units of inverse length. The first term in Eq. (1) is a short-range interaction which decays rapidly on a length scale of $\sim 2/\mu$, and the second term is the long-range "background" interaction.[25] For a pure density functional (i.e. BLYP or BOP) which does not already include a fraction of nonlocal HF exchange, the exchange-correlation energy according to the LC scheme is

$$E_{xc} = E_{c,\text{DFT}} + E_{x,\text{DFT}}^{\text{SR}} + E_{x,\text{HF}}^{\text{LR}}, \qquad (2)$$

where $E_{c,\text{DFT}}$ is the DFT correlational functional, $E_{x,\text{DFT}}^{\text{SR}}$ is the short-range DFT exchange functional, and $E_{x,\text{HF}}^{\text{LR}}$ is the HF contribution to exchange computed with the long-range part of the Coulomb operator. The modified $E_{x,\text{HF}}^{\text{LR}}$ term can be analytically evaluated with Gaussian basis functions,[49] and the short-range $E_{x,\text{DFT}}^{\text{SR}}$ contribution is computed with a modified exchange kernel specific for each generalized gradient approximation (GGA). For the BLYP exchange-correlation functional used in this work, the short-range part of the exchange energy takes the form

$$E_{x,\text{DFT}}^{\text{SR}} = -\frac{1}{2}\sum_\sigma \int \rho_\sigma^{4/3} K_\sigma \left\{1 - \frac{8}{3}a_\sigma\left[\sqrt{\pi}\,\text{erf}\left(\frac{1}{2a_\sigma}\right) + 2a_\sigma(b_\sigma - c_\sigma)\right]\right\}d^3r, \qquad (3)$$

where $\rho_\sigma$ is the density of $\sigma$-spin electrons, and $K_\sigma$ is the GGA part of the exchange functional. The expressions for $a_\sigma$, $b_\sigma$, and $c_\sigma$ are given by

$$a_\sigma = \frac{\mu K_\sigma^{1/2}}{6\sqrt{\pi}\rho_\sigma^{1/3}}, \qquad (4)$$

$$b_\sigma = \exp\left(-\frac{1}{4a_\sigma^2}\right) - 1, \qquad (5)$$

and

$$c_\sigma = 2a_\sigma^2 b_\sigma + \frac{1}{2}. \tag{6}$$

The correlation contribution represented by $E_{c,\text{DFT}}$ in Eq. (2) is left unmodified from its original DFT definition.

The key improvement in the LC scheme is that the short-range DFT exchange interaction is incorporated in the second term of Eq. (2), but the correct long-range interaction is described via HF exchange through the parameter $\mu$. Specifically, the exchange potentials of conventional density functionals have the wrong asymptotic behavior, but the LC scheme ensures that the exchange potential smoothly recovers the exact $-1/r$ dependence at large interelectronic distance. The damping parameter $\mu$ determines the relative contributions of DFT and HF to the exchange-correlation energy. For $\mu = 0$, the LC scheme reduces Eq. (2) to $E_{xc} = E_{c,\text{DFT}} + E_{x,\text{DFT}}$, and all electronic interactions are described with a pure exchange-correlation density functional. Conversely, the $\mu \rightarrow \infty$ limit corresponds to an exchange-correlation functional of the form $E_{xc} = E_{c,\text{DFT}} + E_{x,\text{HF}}$ where all DFT exchange has been replaced by nonlocal HF exchange. The resulting LC scheme is therefore more general than conventional hybrid functionals which are defined with a fixed fraction of nonlocal HF exchange (i.e. B3LYP or PBE0). That is, conventional hybrids incorporate a constant admixture of HF exchange while the LC formalism mixes exchange energy densities based on interelectronic distances at each point in space.

For the dye-sensitized solar cells in this work, we investigated the performance of a long-range-corrected BLYP functional (LC-BLYP) according to the conventional LC scheme described previously. All ground-state geometry optimizations of the coumarin dyes (Fig. 1 and 3) were calculated at the LC-BLYP/6-31G(d,p) level with the damping parameter $\mu$ set to the 0.33 Bohr$^{-1}$ value recommended by Iikura et al.[25] (we also carried out separate optimizations at the B3LYP/6-31G(d,p) level further described in Sec. III. B. and found that the B3LYP relaxed geometries were nearly identical). Geometry optimizations were calculated without symmetry constraints, and root-mean-squared forces converged to within 0.00003 atomic units. In the TDDFT calculations, the lowest singlet-singlet vertical excitation was calculated at the optimized ground-state geometries using a larger 6-31+G(d,p) basis set. The vertical excitation energies for all 8 dyes were also calculated as a function of $\mu$ ranging from 0 to 0.90 Bohr$^{-1}$, in increments of 0.05 Bohr$^{-1}$. For both the ground-state and TDDFT calculations, we used a high accuracy Lebedev grid consisting of 96 radial and 302 angular quadrature points. The LC-BLYP $S_1$ electric dipole moments were evaluated using analytical LC-TDDFT energy derivatives recently implemented by Chiba et al.[26] All *ab initio* calculations were performed with a development version of GAMESS.[50]

## III. Results and Discussion

### A. Excitation energy benchmarks

In each of the ground-state optimizations for the coumarin dyes, the cyanoacrylic acid group becomes coplanar with the polymethine/thiophene chain, indicating a strong conjugation across the valence $\pi$ orbitals of the C=C double bonds. For the NKX-2388, -2311, and -2586 series of dyes, two different stable isomers are possible. As shown in Fig. 3, the configuration at the single bond between the coumarin moiety and the methine chain can have either an *s-cis* or *s-trans* arrangement. In a previous study, it was suggested that the *s-cis* isomer was more stable than the *s-trans* isomer due to steric repulsion between the C=O group in the coumarin and the C≡N group of the cyanoacrylic acid.[12] The LC-BLYP/6-31+G(d,p) calculations support this hypothesis, with the *s-cis* form of the NKX-2388 and -2311 dyes being more stable than the *s-trans* form by 6.1 and 0.2 kcal/mol, respectively. In contrast, steric repulsion becomes negligible for longer polymethine chains, and the trend reverses with the *s-trans* isomer of the NKX-2586 dye being more stable than the *s-cis* isomer by 0.9 kcal/mol.

To investigate the acceptable range of the damping parameter $\mu$, we computed the lowest singlet-singlet vertical excitation energy as a function of $\mu$ for each of the solar cell dyes (including both *cis*- and *trans*-isomers for the NKX-2388, -2311, and -2586 coumarins). In all 8 dyes, this $S_0 \rightarrow S_1$ excitation corresponds to a charge-transfer transition from an orbital delocalized on the coumarin moiety to a localized orbital on the cyanoacrylic acid group (cf. Fig. 2). The ground-to-excited state difference in electric dipole moments, discussed further in Sec. III. B., is also a direct reflection of electronic charge flow associated with this absorption process. Figs. 3 (a)-(e) display the excitation energies of the coumarin dyes compared against recent CC2 calculations by Kurashige et al.[20] The CC2 calculations in Ref. 20 were calculated with an SV(P) basis set which was the largest affordable basis for the longest NKX-2677 dye. The horizontal lines in each figure represent the CC2 excitation energies, and the curved lines denote the TDDFT LC-BLYP/6-31+G(d,p) calculations. For the NKX-2388, -2311, and -2586 coumarins which can exist as *cis*- and *trans*-isomers, solid lines denote excitation energies for the *trans*-isomer while the dashed lines represent the *cis*-isomer.

In all 8 coumarin dyes, the LC-BLYP excitation energies exhibit a very strong $\mu$-dependence which can vary as much as 1.2 eV within the $0 \leq \mu \leq 0.4$ Bohr$^{-1}$ range. It is also interesting to point out that the CC2 lines intersect their corresponding LC-BLYP curves near $\mu = 0.2$ Bohr$^{-1}$ for all of the coumarin dyes. Using the CC2 energies as reference values, Fig. 5 shows the total root-mean-square error (RMSE) in the excitation energies for all 8 dyes as a function of $\mu$. As seen in Fig. 5, the RMSE curve has a deep minimum at $\mu = 0.17$ Bohr$^{-1}$ with an RMS error of 0.04 eV. Perhaps surprisingly, this value of $\mu$ is considerably smaller than the standard 0.33 Bohr$^{-1}$ value recommended by Iikura et al.[25] In a very recent study by Rohrdanz and Herbert,[43] it was found that excited-state properties were strongly $\mu$-dependent and were significantly different for small and large molecules. In particular, it was suggested that smaller values of $\mu$ may be more appropriate for excitation energies in large molecules since a smaller value of $\mu$ enables the short-range Coulomb operator to fully decay to zero on the length scale of the molecule. The present $\mu = 0.17$ Bohr$^{-1}$ result supports their suggestion and is comparable to the $\mu = 0.18$ and 0.20 Bohr$^{-1}$ values they obtained for valence excitations in anthracene and pyridazine respectively.

**B. Overall trends and properties**

In this section we consider the overall general trends in excitation energies and properties for the coumarin series of dyes. To assess the performance of the LC-BLYP approach, we also carried out separate geometry optimizations and excitation energies using the popular B3LYP functional which incorporates a fixed combination of 20% Hartree-Fock, Slater, and Becke's GGA correction in the exchange contribution. All B3LYP ground-state geometry optimizations were computed using the 6-31G(d,p) basis set, and vertical excitation energies were calculated at these optimized geometries using a larger 6-31+G(d,p) basis set. The fully relaxed B3LYP geometries were nearly identical to the LC-BLYP geometries and had slightly longer C=C double bonds by only 0.02 Å. Table I lists excited-state energies and properties for the LC-BLYP functional with the range parameter $\mu$ set to 0.17 Bohr$^{-1}$, and Table II gives the corresponding results for the B3LYP functional.

On a qualitative level, both functionals reproduce the expected trend that the excitation energy decreases as the $\pi$-conjugated chain becomes longer. Fig. 6 shows in more detail the general trend in the $S_0 \rightarrow S_1$ transition energies between the LC-BLYP, B3LYP, and CC2 results for all 8 solar cell dyes. The diagonal line in Fig. 6 represents an ideal 100% agreement between the CC2 energies and the corresponding TDDFT results. The close agreement between the LC-BLYP and CC2 results is of course not surprising considering that we explicitly minimized the RMS error for all 8 dyes by re-optimizing the range parameter $\mu$. Instead, the most significant feature of Fig. 6 is the high degree of linear correlation exhibited by LC-BLYP, indicative of a simple, systematic error in these excitation energies. Using a simple linear fit to the TDDFT data points, one obtains a high statistical correlation with $R^2 = 0.98$ for the LC-BLYP results. Somewhat surprisingly, it was found that even if $\mu$ was increased up to 0.33 Bohr$^{-1}$ for all dyes (which severely overestimates the excitation energies), the LC-BLYP energies still showed a high statistical correlation with $R^2 = 0.98$ (in other words, increasing the value of $\mu$ only shifts all the excitation energies upwards by the same amount and does not change the relative energy ordering between each of the coumarin dyes).

In contrast, the same linear fitting procedure for the B3LYP excitation energies yields a poorer correlation with an $R^2$ value of 0.85. The B3LYP calculations show good agreement with the CC2 results for the small C343 and NKX-2388 dyes, but the trend in excitation energies is poorly described for the larger coumarin dyes. One can (unjustifiably) attempt to "correct" the B3LYP results by minimizing the RMS error with the CC2 energies and re-optimizing the fraction of exact exchange in the functional. Increasing the percentage of exchange to 100% in B3LYP does significantly improve the linear fitting $R^2$ value; however, this empirical procedure also ruins the balance between exchange and correlation errors, and the final excitation energies become severely overestimated. In particular, we found that it was not possible to simultaneously obtain both accurate excitation energies and reasonable $R^2$ values by adjusting the fraction of exact exchange in B3LYP for the present coumarin dyes.

To gain further insight into the electronic structure of the excited charge-transfer states, we also computed the $S_0/S_1$ electric dipole moments and oscillator strengths of the $S_0 \rightarrow S_1$ transitions. In all 8 solar cell dyes, the $S_0$ molecular orbital is delocalized throughout the entire molecule with maximum density from the nitrogen lone-pair and the $\pi$ orbitals of the surrounding aryl substituents. The $S_1$ molecular orbital, on the other hand, is primarily a $\pi^*$ orbital which is localized on the conjugated chain and cyanoacrylic acid units with density concentrated on the cyano- and carboxylic groups (cf. Fig. 2). As anticipated from this increased orbital localization, the $S_1$ electric dipole moment is considerably larger than the $S_0$ dipole, an electronic signature which reflects a sizable charge transfer in the excited state. Although this charge-transfer effect is captured in both functionals, the B3LYP excited-state dipoles for the larger NKX-2586 and -2677 dyes are significantly overestimated relative to the LC-BLYP results. According to the simple two-level system model by Oudar and Chemla,[51] an increase in the excited-state dipole moment concomitantly reduces the corresponding $S_0 \rightarrow S_1$ oscillator strength in the same molecule. This tendency is in accord with the B3LYP results which severely underestimate the CC2 oscillator strengths for the NKX-2586 and -2677 dyes. In contrast, the LC-BLYP calculations predict a smaller change in the dipole moment, and therefore yield larger oscillator strengths in better agreement with the CC2 calculations.

Again, it is interesting to see the effect of increasing the coefficient of exact exchange on excited-state properties for the coumarin series of dyes. Since B3LYP incorporates a fixed fraction of 20% HF exchange, the exchange potential has a –0.2/$r$ dependence at large interelectronic distances. As a result, one can expect that this incorrect exchange potential is not attractive enough, leading to an overestimation of electron transfer and hence a larger dipole moment. Indeed, increasing the percentage of exchange to 100% in B3LYP does decrease the magnitude of the $S_1$ dipole moment towards the correct value but, at the same time, also overestimates the vertical excitation energies. Although a modified B3LYP functional with 100% exchange recovers the asymptotically correct –1/$r$ behavior of the exchange potential, this modification overcompensates for the error in the DFT correlation part. On the other hand, the LC technique offers significantly more flexibility compared to conventional DFT hybrids which incorporate a constant admixture of exchange. By optimizing the range parameter $\mu$, one can adjust the length scale where short-range exchange-correlation effects dominate while still retaining the correct asymptotic behavior of the exchange potential. While we have described excited-state energies and properties of several coumarin solar cell dyes, it is clear that the application and limits of the LC-TDDFT treatment still require more study. Nevertheless, the resulting excited-state trends and properties indicate that long-range exchange interactions do indeed play a significant role in these systems, and the LC-TDDFT technique offers an improved treatment over conventional DFT hybrids for capturing the excited-state dynamics in coumarin solar cell dyes.

**IV. Conclusion**

In this study, we have theoretically investigated the excited-state energies and properties in a series of coumarin dye molecules for use in solar cell applications. For each of the 8 coumarin dyes, excited-state energies and properties were obtained using the TDDFT formalism in conjunction with a functional modified specifically for long-range intramolecular charge-transfer. In all cases, we find that the TDDFT valence excitation energies show a very strong dependence on the range-separation parameter, $\mu$. To investigate the acceptable range of $\mu$, an extensive comparison was made between LC-BLYP excitation energies and previous CC2 calculations. In agreement with existing benchmark studies, we find that the RMS error in the LC-BLYP functional is minimized for a value of $\mu$ which is significantly smaller than the recommended value used in small molecular systems. Using this re-optimized value of $\mu$, we find that the long-range-corrected LC-BLYP functional significantly improves the poor description given by B3LYP and correctly reproduces the correct trends among the larger coumarin dyes. In particular, a simple statistical analysis demonstrates that the LC-TDDFT treatment provides a more consistent picture of excitation energies as a function of DSC molecular size.

Among the DSCs studied here, we also calculated a large increase in the $S_1$ electric dipole moment with respect to that of the ground state, indicating a sizable charge transfer associated with the $S_0 \rightarrow S_1$ excitation. The amount of charge-transfer involved in this electronic transition is significantly overestimated by B3LYP, leading to large dipole moments and underestimated oscillator strengths in the NKX-2586 and NKX-2677 dyes. We also note that empirically increasing the percentage of exchange in B3LYP does improve certain properties such as $S_1$ dipoles and $R^2$ correlation values, but the same procedure also corrupts the balance between exchange and correlation errors with the final excitation energies becoming severely overestimated. Although the LC treatment still partially relies on a HF exchange/DFT correlation error cancellation, this method offers the "best of both worlds" in the sense that one can adjust $\mu$ so that short-range exchange-correlation effects dominate while still retaining the physically-correct asymptotic behavior of the exchange potential.

In closing, the present study provides indications that long-range exchange corrections play a vital role in describing the excited-state dynamics in solar cell dyes. For a properly-optimized value of the range parameter $\mu$, the LC-TDDFT technique offers an improved treatment of excited-state properties and overall trends compared to conventional hybrid functionals like B3LYP. Although the present study has focused on several coumarin solar cell dyes where reliable CC2 data is available, further tests on large systems are still needed to understand the limitations of the LC approach. In particular, we are hopeful that the LC-TDDFT formalism will help verify and predict the excited-state properties of light-harvesting organic systems which we are currently synthesizing. With this in mind, we anticipate that the understanding of critical features such as excitation energies and charge-transfer states using the LC-TDDFT technique will provide a step towards this goal.

**Acknowledgements**

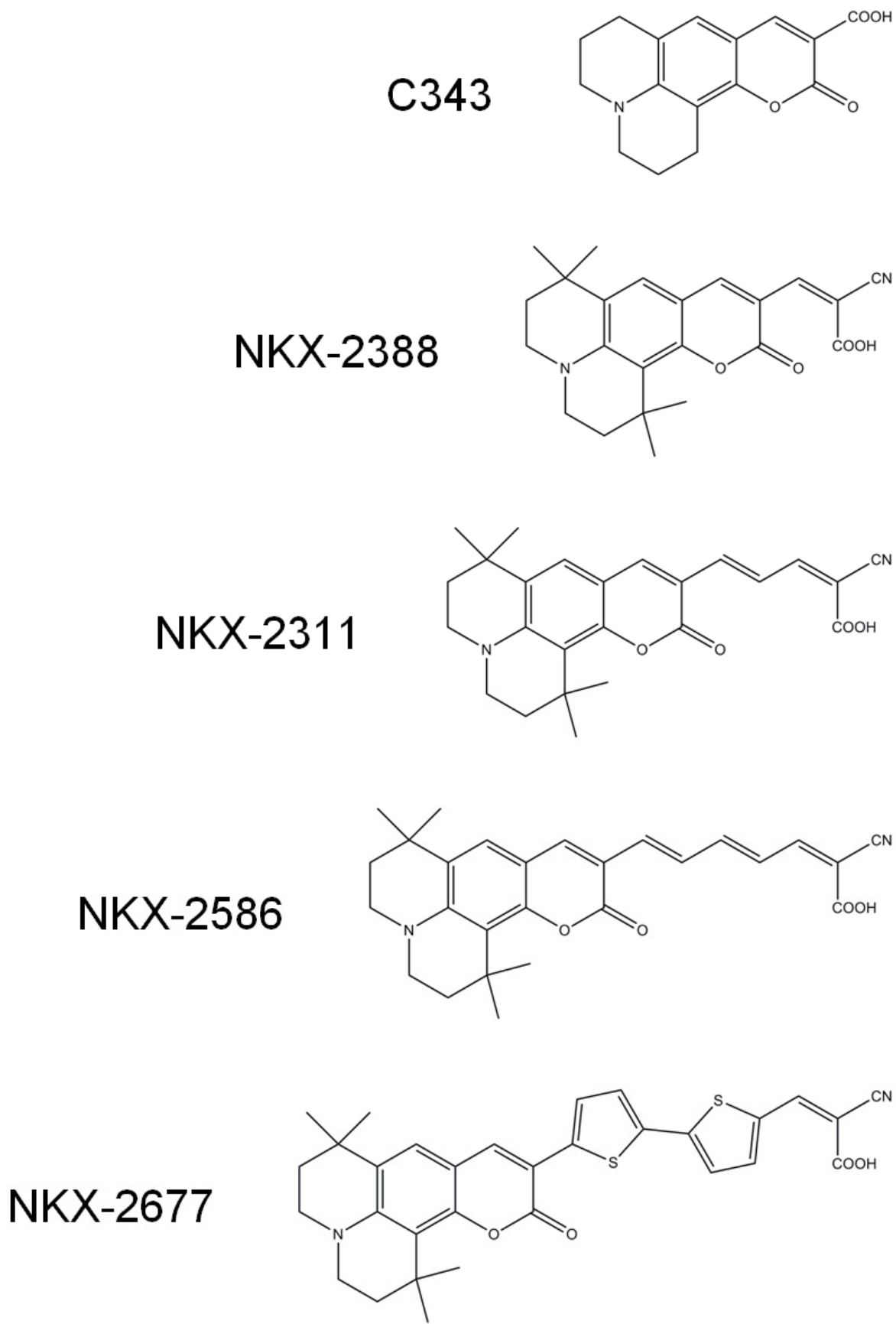

**FIG. 1.** Molecular structures of C343, NKX-2388, NKX-2311, NKX-2586, and NKX-2677. Each dye consists of an electron-donating aniline group and an electron-accepting cyanoacrylic acid anchoring group. The polymethine/thiophene units function as an electron conductor by providing a strong $\pi$-conjugation across the donor and anchoring groups.

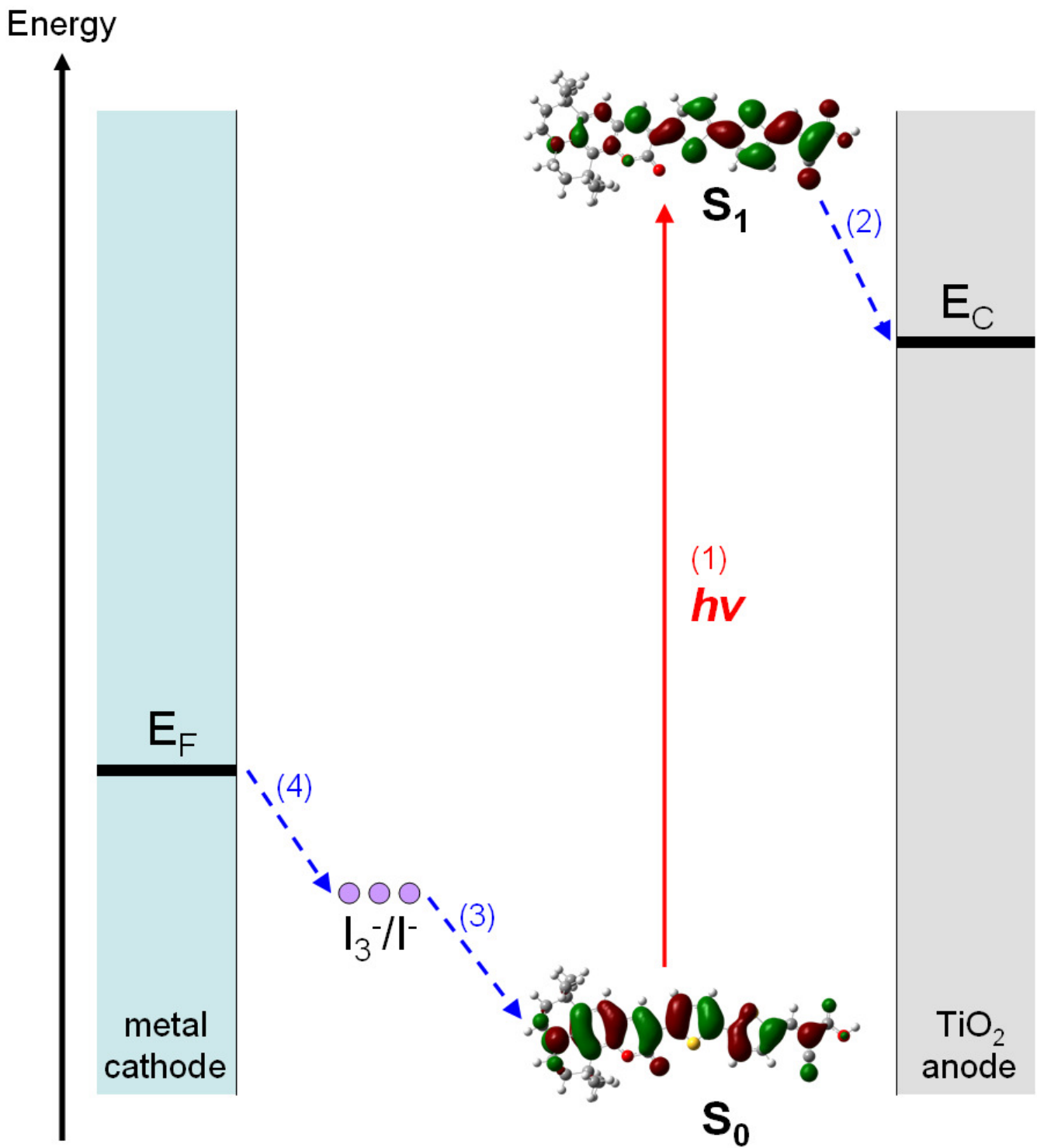

**FIG. 2.** Simplified schematic of the energetic processes occurring in a conventional dye-sensitized solar cell: (1) absorption of a photon promotes an electron into a $S_1$ charge-transfer state, (2) the excited electron migrates into the energetically-lower $TiO_2$ anode, (3) the oxidized DSC is regenerated with electrons from the energetically-higher $I_3^-$ electrolyte, and (4) the $I_3^-$ ions are reduced to I- ions by the metal cathode, thus completing the cycle.

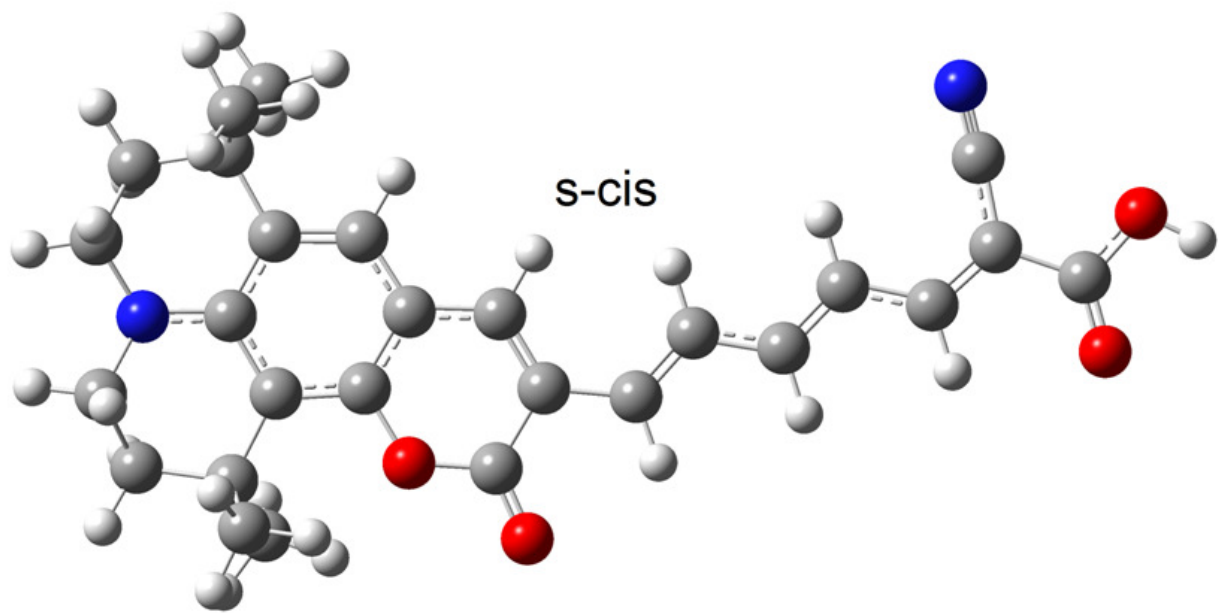

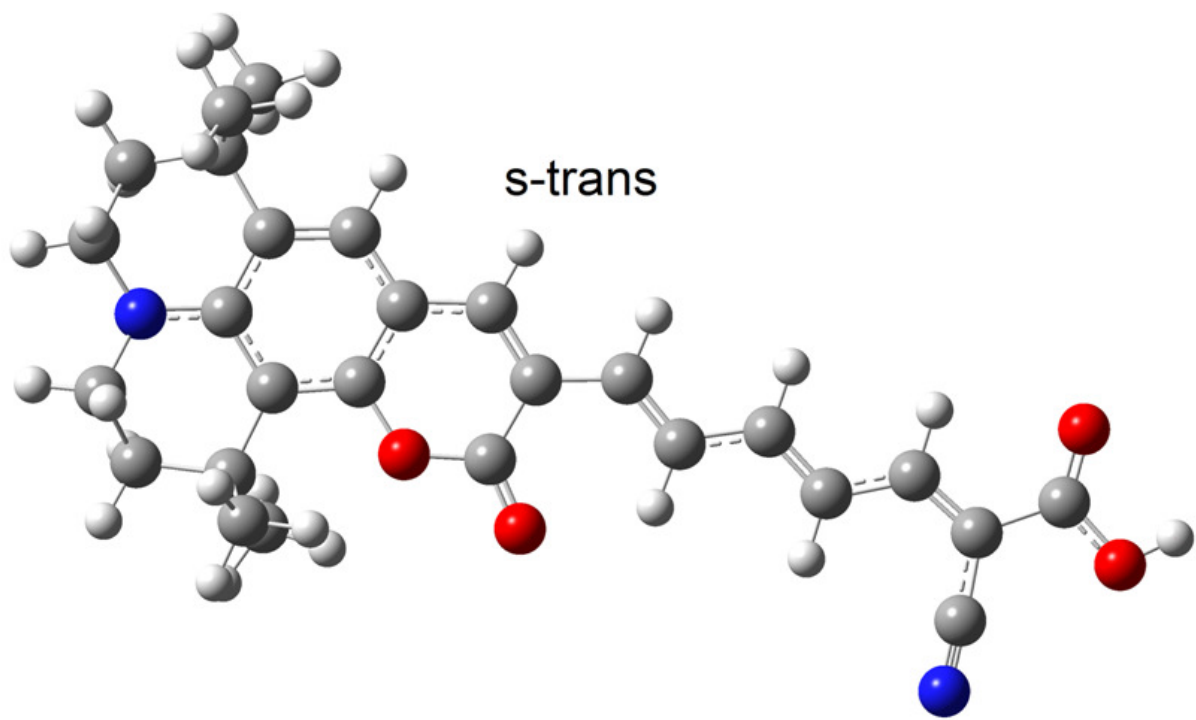

**FIG. 3.** Optimized molecular structures of the *s-cis* and *s-trans* isomers of NKX-2311 at the LC-BLYP/6-31G(d,p) level of theory.

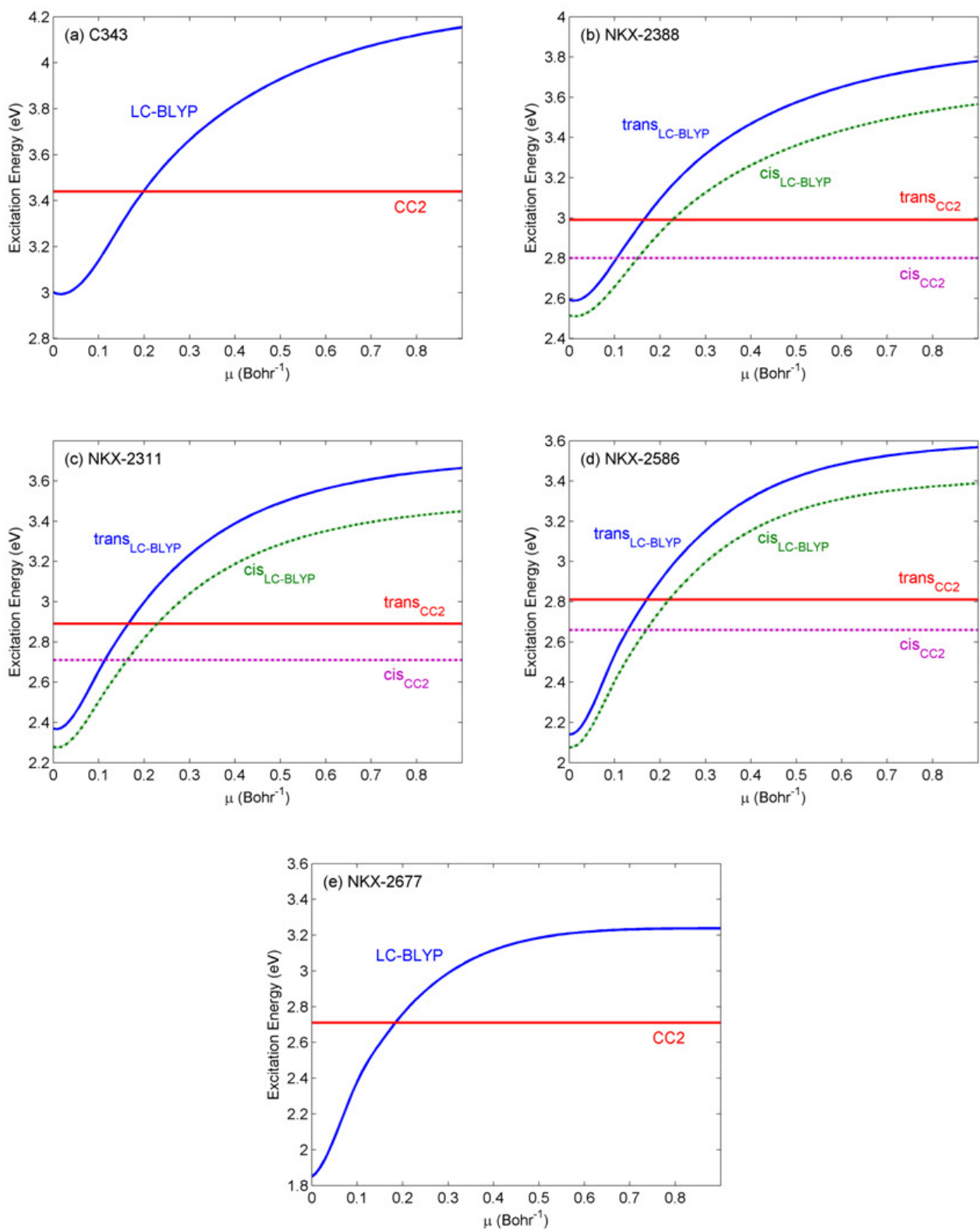

**FIG. 4.** Vertical excitation energies as a function of the range parameter $\mu$ for (a) C343, (b) *s-trans*-NKX-2388 and *s-cis*-NKX-2388, (c) *s-trans*-NKX-2311 and *s-cis*-NKX-2311, (d) *s-trans*-NKX-2586 and *s-cis*-NKX-2586, and (e) NKX-2677. The horizontal lines in each figure represent the CC2 excitation energies, and the curved lines denote the TDDFT LC-BLYP calculations. The solid lines in Figs. 4. (b)-(d) denote excitation energies for the *trans*-isomer, while dashed lines represent the *cis*-isomer.

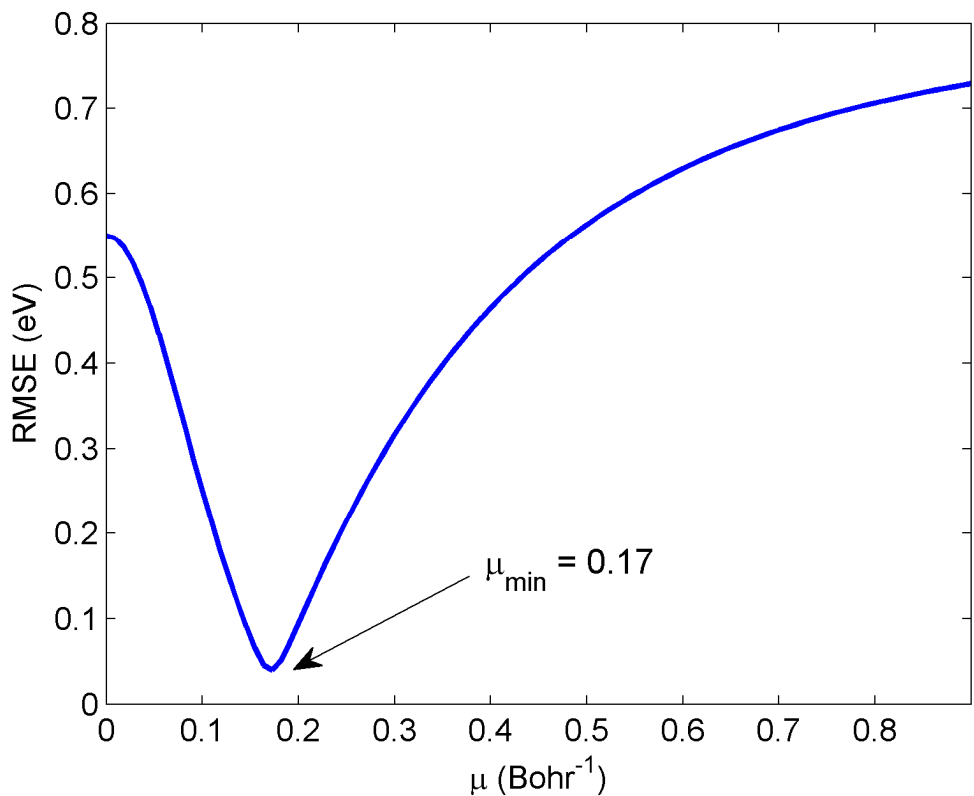

**FIG. 5.** Total root-mean-square error (RMSE) in the excitation energies for all 8 coumarin dyes as a function of $\mu$. The RMSE curve has a deep global minimum at $\mu = 0.17$ with an RMS error of 0.04 eV.

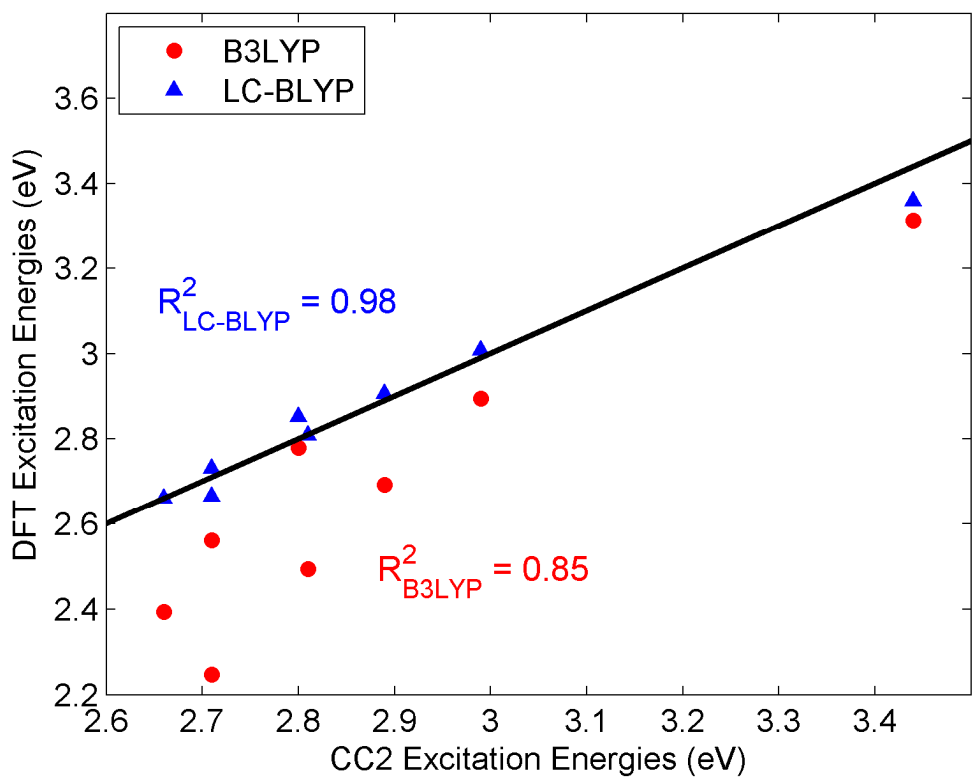

**FIG. 6.** Comparison between TDDFT and CC2 excitation energies for all 8 coumarin dyes. The diagonal line represents a perfect match between CC2 and TDDFT energies. The $R^2$ values for the LC-BLYP and B3LYP results were obtained from a simple linear fit to the data points themselves and not calculated with respect to the diagonal line shown in the figure.

**TABLE 1:** TDDFT $S_0 \rightarrow S_1$ excitation energies, oscillator strengths, and dipole moments of courmarin dyes computed at the LC-BLYP/6-31+G(d,p) level of theory with the range parameter $\mu = 0.17$ Bohr$^{-1}$.

| LC-BLYP | TDDFT Absorbance (eV) | CC2 Absorbance (eV)[a] | TDDFT Oscillator Strength | CC2 Oscillator Strength[a] | $S_0$ Dipole (Debye) | $S_1$ Dipole (Debye) |
|---|---|---|---|---|---|---|
| C343 | 3.36 | 3.44 | 0.573 | 0.738 | 9.78 | 13.54 |
| NKX-2388 (s-trans) | 3.01 | 2.99 | 0.877 | 1.064 | 11.29 | 16.22 |
| NKX-2388 (s-cis) | 2.85 | 2.80 | 0.805 | 1.004 | 9.10 | 14.10 |
| NKX-2311 (s-trans) | 2.91 | 2.89 | 1.337 | 1.511 | 12.49 | 19.08 |
| NKX-2311 (s-cis) | 2.73 | 2.71 | 1.117 | 1.329 | 9.49 | 16.34 |
| NKX-2586 (s-trans) | 2.81 | 2.81 | 1.828 | 2.007 | 13.26 | 21.41 |
| NKX-2586 (s-cis) | 2.66 | 2.66 | 1.519 | 1.740 | 10.03 | 18.48 |
| NKX-2677 | 2.67 | 2.71 | 1.763 | 2.174 | 13.11 | 19.48 |

[a]Excitation energies and oscillator strengths from Ref. 20

**TABLE 2:** TDDFT $S_0 \rightarrow S_1$ excitation energies, oscillator strengths, and dipole moments of coumarin dyes computed at the B3LYP/6-31+G(d,p) level of theory.

| B3LYP | TDDFT Absorbance (eV) | CC2 Absorbance (eV)[a] | TDDFT Oscillator Strength | CC2 Oscillator Strength[a] | $S_0$ Dipole (Debye) | $S_1$ Dipole (Debye) |
|---|---|---|---|---|---|---|
| C343 | 3.31 | 3.44 | 0.602 | 0.738 | 10.39 | 13.96 |
| NKX-2388 (s-trans) | 2.89 | 2.99 | 0.934 | 1.064 | 12.46 | 16.92 |
| NKX-2388 (s-cis) | 2.78 | 2.80 | 0.869 | 1.004 | 10.52 | 14.44 |
| NKX-2311 (s-trans) | 2.69 | 2.89 | 1.350 | 1.511 | 14.12 | 20.65 |
| NKX-2311 (s-cis) | 2.56 | 2.71 | 1.188 | 1.329 | 11.45 | 17.14 |
| NKX-2586 (s-trans) | 2.49 | 2.81 | 1.703 | 2.007 | 15.40 | 24.68 |
| NKX-2586 (s-cis) | 2.39 | 2.66 | 1.544 | 1.740 | 12.54 | 20.59 |
| NKX-2677 | 2.25 | 2.71 | 1.508 | 2.174 | 15.59 | 28.70 |

[a]Excitation energies and oscillator strengths from Ref. 20